\documentclass[aps,prb,twocolumn,superscriptaddress]{revtex4-1}

\usepackage{amssymb}
\usepackage{graphicx}
\usepackage{amsmath}

\usepackage{hyperref}

\begin{document}

\title{Out-of-Time-Order Correlation at a Quantum Phase Transition}
\author{Huitao Shen}
\affiliation{Institute for Advanced Study, Tsinghua University, Beijing, 100084, China}
\affiliation{Department of Physics, Massachusetts Institute of Technology, Cambridge, MA 02139, USA}
\author{Pengfei Zhang}
\affiliation{Institute for Advanced Study, Tsinghua University, Beijing, 100084, China}
\author{Ruihua Fan}
\affiliation{Institute for Advanced Study, Tsinghua University, Beijing, 100084, China}
\affiliation{Department of Physics, Peking University, Beijing, 100871, China}
\author{Hui Zhai}
\affiliation{Institute for Advanced Study, Tsinghua University, Beijing, 100084, China}
\date{\today}

\begin{abstract}
Motivated by the recent studies of out-of-time-order correlation functions and the holographic duality, we propose the QCP Conjecture, which is stated as: For a many-body quantum system with a quantum phase transition, the Lyapunov exponent extracted from the out-of-time-order correlators will exhibit a maximum around the quantum critical region. We first demonstrate that the Lyapunov exponent is well-defined in one-dimensional Bose-Hubbard model with the help of the OTOC-RE theorem. We then support the conjecture by numerically computing the out-of-time-order correlators. We also compute the butterfly velocity, and propose an experiment protocol of measuring this correlator without inverting the Hamiltonian. 
\end{abstract}

\maketitle

\section{Introduction}

Recently there is an increasing interest in the out-of-time-order correlation functions (OTOC) \cite{Larkin1969,Shenker2014,Shenker2013,Almheiri2013,Shenker2014a,Kitaev2014,Roberts2015,Roberts2015a,Maldacena2016,Stanford2016,Hosur2016,Gu2016,Kitaev2015,Maldacena2016a,Jevicki2016,Jensen2016,Maldacena2016b,Engelsoy2016,Fu2016,Danshita2016,Swingle2016,Zhu2016,Yao2016} defined as
\begin{equation}
F(t)=\langle \hat{W}^\dagger(t)\hat{V}^\dagger(0)\hat{W}(t)\hat{V}(0)\rangle_\beta,
\label{otoc}
\end{equation}
where $\hat{W}$ and $\hat{V}$ are normally chosen as local operators. $\hat{W}(t)\equiv e^{i\hat{H}t}\hat{W}e^{-i\hat{H}t}$ , and $\langle \ldots\rangle_\beta\equiv\mathrm{tr}[e^{-\beta H}\ldots]$ denotes the thermal average at temperature $1/\beta=k_\text{B}T$. Intuitively, this correlation function can be considered as the overlap of two states $\langle y|x\rangle$, where $|x\rangle=\hat{W}(t)\hat{V}(0)|\beta\rangle$ and $|y\rangle=\hat{V}(0)\hat{W}(t)|\beta\rangle$. $ |\beta\rangle\equiv\sum_n e^{-\beta E_n/2}/\sqrt{Z}|n\rangle|\tilde{n}\rangle $ is the thermofield double state \cite{Israel1976}. $ Z=\mathrm{tr}\,e^{-\beta H} $ is the partition function, $ |n\rangle  $ and $ |\tilde{n}\rangle $ are the corresponding energy eigenstates of the Hamiltonian but in different Hilbert spaces. In this sense, the inner product $\langle y|x\rangle$ measures the difference in the outcome when the order of two operations $\hat{V}(0)$ and $\hat{W}(t)$ is exchanged. The exponential deviation of the normalized OTOC
\begin{equation}
\tilde{F}(t)=\frac{\langle y|x\rangle}{\sqrt{\langle x|x\rangle\langle y|y\rangle}},
\end{equation}
from unity diagnoses the chaos and the so-called ``butterfly effect" in a quantum many-body system \cite{Shenker2014,Shenker2013,Almheiri2013,Shenker2014a,Kitaev2014,Roberts2015,Roberts2015a,Maldacena2016,Stanford2016,Hosur2016,Gu2016}. This deviation can be explicitly written as $\tilde{F}(t)=\alpha_0-\alpha_1 e^{\lambda_\mathrm{L}(t-t_0)}$ ($\alpha_0\approx 1$). Here the deviation starts from $t_0$, and $\lambda_\mathrm{L}$ defines the Lyapunov exponent for this quantum system. 

\begin{figure}[t]
\includegraphics[width=\columnwidth]
{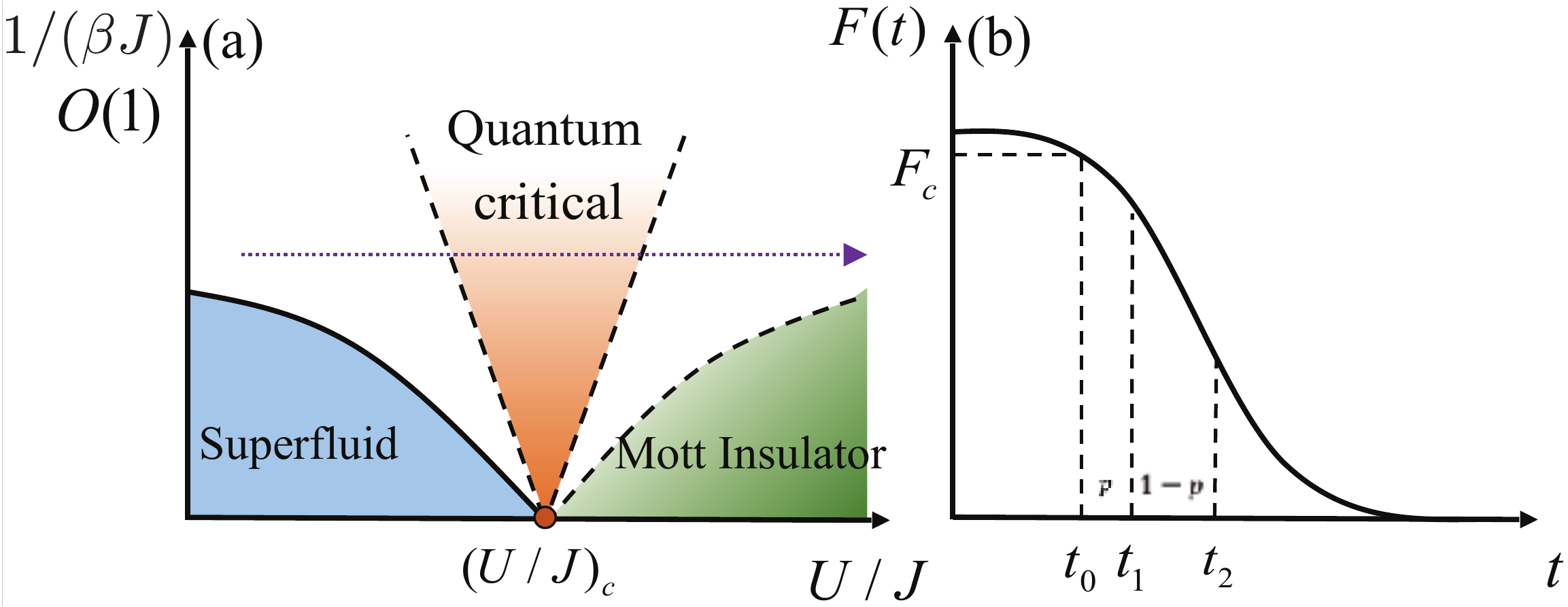}
\caption{(a) Schematic phase diagram of the Bose-Hubbard model. The dotted line illustrates the parameter regime that is considered in this work. (b) Schematic OTOC and the fitting scheme to obtain the Lyapunov exponent. See Sec.~\ref{lya} for more details. }
\label{schematic}
\end{figure}

It turns out that the same correlator has emerged in the gravity physics, in the context of which it describes a bulk scattering near the horizon and characterizes the information scrambling \cite{Shenker2014,Shenker2013,Almheiri2013,Shenker2014a,Kitaev2014}. More interestingly, it is shown recently that for quantum systems, the Lyapunov exponent is always bounded by $2\pi/\beta$ \cite{Maldacena2016}. If a quantum many-body system has an exact holographic duality to a black hole at finite temperature \cite{Maldacena2000,Witten1998,Gubser1998}, the Lyapunov exponent will saturate the bound $\lambda_\mathrm{L}=2\pi/\beta$. While a more nontrivial speculation is that if the Lyapunov exponent of a quantum system saturates this bound, this system displays a holographic duality to a black hole \cite{Maldacena2016}. In this sense, the previously defined Lyapunov exponent measures how close a quantum many-body system is to have a holographic duality to a black hole. A quantum mechanical model, which is known as the ``Sachdev-Ye-Kitaev" model \cite{Sachdev1993,Kitaev2015}, has been shown to have the emergent conformal symmetry \cite{Sachdev1993,Parcollet1998,Kitaev2015,Maldacena2016a} and the holographic duality \cite{Jevicki2016,Jensen2016,Maldacena2016b,Engelsoy2016}. The OTOC in this model can be calculated explicitly and the Lyapunov exponent is found to saturate the bound \cite{Kitaev2015,Maldacena2016a,Fu2016}. 

In this work we are interested in studying the OTOC for more realistic models. We will mainly focus on the Bose-Hubbard model (BHM). This model has been well-studied as a textbook example for quantum phase transitions \cite{Fisher1989,Sachdev2011}. Since its first realization in the optical lattice in 2011, the BHM has become one of the most well-studied models experimentally in cold atom physics \cite{Jaksch2005,Bloch2008,Yukalov2009}. The Hamiltonian of the BHM is
\begin{equation}
\hat{H}=-J\sum\limits_{\langle ij\rangle}(\hat{b}^\dag_i\hat{b}_j+\mathrm{h.c.})+\frac{U}{2}\sum\limits_{i}\hat{n}_i(\hat{n}_i-1),
\end{equation}
where $\hat{b}_i$ is the spinless boson operator at $i$-th site and $\hat{n}_i=\hat{b}^\dag_i\hat{b}_i$ is the boson number operator. 
At integer filling, as $U/J$ increases, this model exhibits a quantum phase transition from the superfluid phase to the Mott insulator phase. Fig.~\ref{schematic}(a) is the schematic phase diagram for the BHM \cite{Kato2008,Sachdev2011}. Since there is also an emergent conformal symmetry near the critical point, and the quantum critical region is so strongly interacting that there are no well-defined single-particle excitations, it is believed that a $(2+1)$-dimensional BHM at the quantum critical regime is dual to a gravity model in the four-dimensional Anti-de Sitter space \cite{Sachdev2012,Fujita2015}. Motivated by this argument, along with the aforementioned insight from the recent studies of the OTOC, we propose a \textit{QCP Conjecture} for the Lyapunov exponent, which is stated as: \textit{the Lyapunov exponent will display a maximum around the quantum critical region}. In the BHM, we will consider increasing $U/J$ across the quantum critical region with a temperature higher than the superfluid transition temperature, as shown by the dotted line in Fig.~\ref{schematic}(a). 

Hereafter we present several calculations to support this conjecture. Due to the lack of a general effective scheme to calculate the OTOC in strongly interacting systems, we perform an exact diagonalization calculation, in which we first obtain all eigenstates for this many-body system and then compute the time-evolution under the basis of these eigenstates. The calculation is limited to a one-dimensional BHM up to 7 sites. Indeed, this is not an ideal model to demonstrate our conjecture. Our results suffer from the finite-size effect, and the original proposal of the holographic duality is for a $(2+1)$-dimensional BHM. Nevertheless, as we will see, the results support our conjecture. 

The paper is organized as follows. In Sec.~\ref{expo} we first demonstrate that the Lyapunov exponent is well-defined in one-dimensional BHM through both numerical and conformal field theory analysis with the help of the OTOC-RE theorem \cite{Fan2017}. In Sec.~\ref{lya}, we then extract the Lyapunov exponents at various parameter regimes to support our conjecture. Since one-dimensional BHM has spatial dimension, we also extract the butterfly velocity in Sec.~\ref{but}. Finally in Sec.~\ref{expp}, we propose experimental protocols to measure the OTOC, making it feasible to test our conjecture in the laboratory. 

\section{Exponential Deviation of the OTOC}
\label{expo}
We first argue that although it is not a fully chaotic model, the OTOC of BHM should deviate exponentially in time. The argument is based on the OTOC-RE theorem in Ref.~\cite{Fan2017}, which relates the OTOC at equilibrium and the second R\'enyi entropy (RE) growth after a local quench. By both numerical calculation and the conformal field theory (CFT) analysis, we show that the second R\'enyi entropy in BHM grows linearly in time after a local quench at finite temperature, implying an exponential deviation of the OTOC. 

The OTOC-RE theorem is stated as follows: Consider an equilibrium system at temperature $T$ described by the density matrix $\hat{\rho}=e^{-\beta\hat{H}}$. When it is quenched by an operator $\hat{O}$ at time $t=0$, the density matrix becomes proportional to $\hat{O}\hat{\rho}\hat{O}^\dag$ and begins to evolve. We then divide the system into two subsystems as $A$ and $B$. The second R\'enyi entropy on $ A $ is defined as $S_A^{(2)}=-\log(\mathrm{tr}[\hat{\rho}_A^2])$, where $ \hat{\rho}_A=\mathrm{tr}_B[\hat{\rho}] $ is the reduced density matrix of $ A $. In Ref.\cite{Fan2017}, we showed that this R\'enyi entropy is related to the summation of modified OTOCs at temperature $T/2$:
\begin{align}
\exp(-S_A^{(2)})=&\sum_{W\in B}\mathrm{tr}\left[\hat{W}^\dagger(t)\hat{O}e^{-\beta \hat{H}}\hat{O}^\dagger \hat{W}(t)\hat{O}e^{-\beta \hat{H}}\hat{O}^\dagger \right] \notag \\
=&\sum_{W\in B}\mathrm{tr}\left[e^{-2\beta H}\hat{W}^\dagger(t-2i\beta)\hat{O}(-2i\beta)\right. \notag\\ 
&\quad\left.\hat{O}^\dagger(-i\beta)\hat{W}(t-i\beta)\hat{O}(-i\beta)\hat{O}^\dagger(0) \right], \label{OTOC_EE}
\end{align}
The summation over $\hat{W}$ is taken over the complete set of operators in system $B$, and $ \hat{V} $ is fixed to be $ \hat{O}\hat{O}^\dagger $. 

Using this theorem, the dynamics of $S_A^{(2)}$ after a quench is related to the behavior of OTOC under the following two conditions, which are assumed to be true: (i) For long time $t\gg \beta$ each term in the R.H.S. of Eq.~\eqref{OTOC_EE} approaches the OTOC $\mathrm{tr}\left[e^{-2\beta \hat{H}}\hat{W}^\dagger(t)\hat{O}\hat{O}^\dagger \hat{W}(t)\hat{O}\hat{O}^\dagger \right]$ \cite{Maldacena2016,Yao2016}; (ii) Different terms in the summation of the R.H.S. of Eq.~\eqref{OTOC_EE} have similar behaviors. In the rest of this section we will show that during certain time interval after a local quench, $S_A^{(2)}$ will increase linearly with time $t$, which further indicates an exponential deviation of OTOC because of Eq.~\eqref{OTOC_EE}.  
\begin{figure}
\includegraphics[width=0.7\columnwidth]{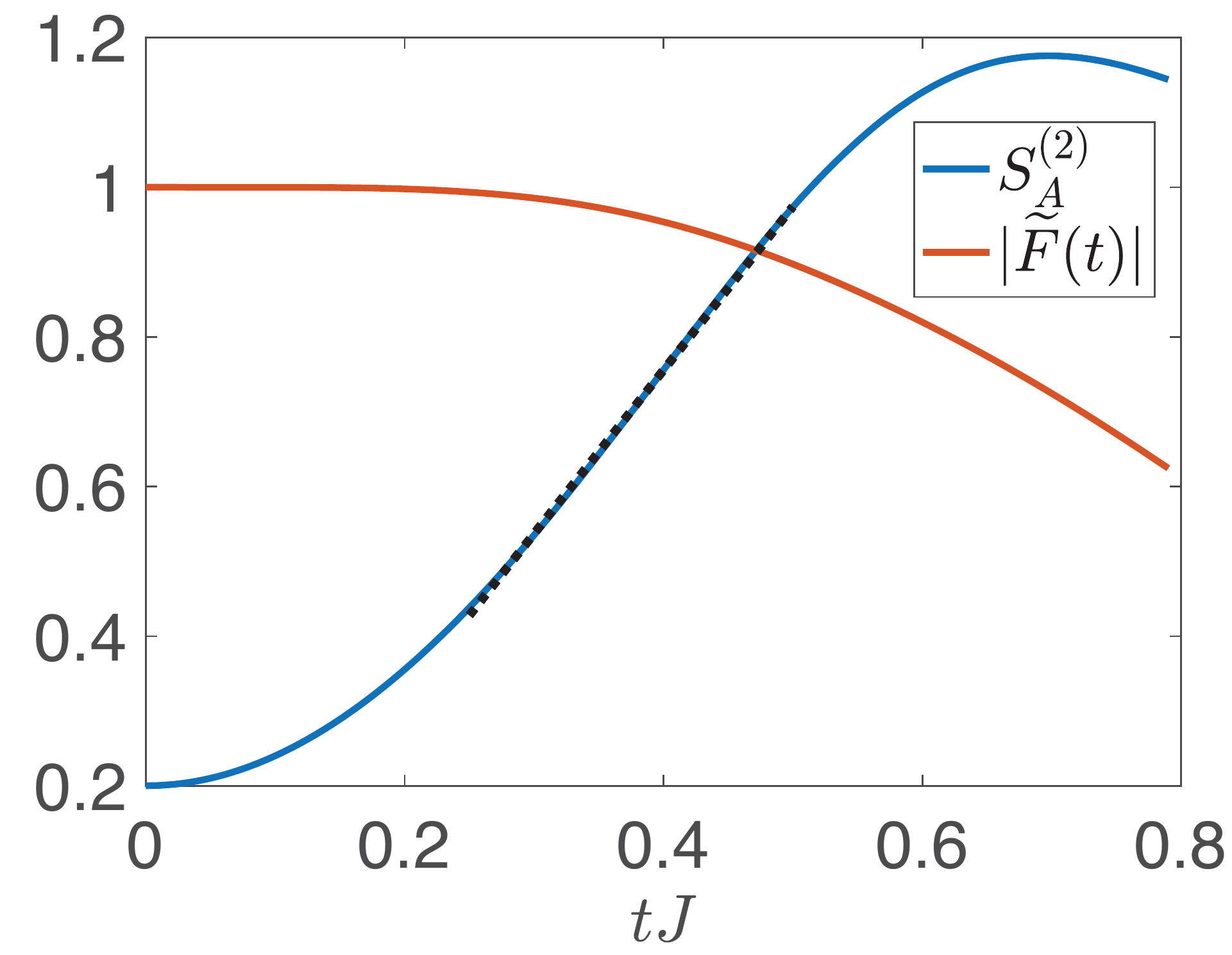}
\caption{The growth of the second R\'enyi entropy $S_A^{(2)}$ and the normalized OTOC $ |\tilde{F}(t)| $ as functions of time $tJ$ for $U/J=10$ at $\beta J=0.9$ and $N=L=6$ with a periodic boundary condition. The linear growth regime of $S_A^{(2)}$ is indicated by a fitted dashed black line. See the main text for more details on the operator choice. }
\label{entropy}
\end{figure}

To study the entropy growth after a local quench for the one-dimensional BHM, we first numerically calculate the second R\'enyi entropy for a 6-site chain using exact diagonalization method. Here we consider a quench that removes a boson at the third site, which corresponds to a quench operator $\hat{b}_3$. Then we divide the system into two equal halves in order to calculate the second R\'enyi entropy. The result is shown in Fig.~\ref{entropy}, where $S_A^{(2)}$ clearly exhibits a linear growth within a time interval. For comparison, we also plot an OTOC with a similar setup by taking $\hat W=\hat b_3$ and $\hat V=\hat b_4$. Clearly, it is during the same time interval that the OTOC starts to deviate from unity. In this sense, the linear growth of the second R\'enyi entropy implies an exponential behavior of the OTOC.   

In fact, specifically for critical one-dimensional model, it is possible to obtain the linear growth of the entropy after a local quench by a CFT analysis. Consider two half-infinite subsystems $A$ and $B$ at equilibrium of temperature $T$. We mimic a local quench by joining the two subsystems into a whole system at $t=0$. The reduced density matrix for subsystem $ A $ at time $t$ is now
\begin{align}
\hat{\rho}_A= \mathrm{tr}_B\left[\exp(-i \hat{H}t)\exp(-\beta \hat{H}^\prime)\exp(i \hat{H}t) \right], \label{cftrho}
\end{align}
where $\hat{H}'$ is the Hamiltonian for separated $A$ and $B$, and $\hat{H}$ is the Hamiltonian for the whole system. Following the treatment in Ref.~\cite{Calabrese2009}, by introducing $n$ replicas and the twist field, one can reduce the problem of computing $ n $-th R\'enyi entropy to the calculation of a single-point correlation function on a manifold with a boundary. More details could be found in the Appendix. The final result is
\begin{align}
S^{(2)}_A=\frac{c}{8}\log(\sinh(\pi T t))+\mathrm{const.}, \label{cftres}
\end{align}
and the long time behavior is given by 
\begin{align}
S^{(2)}_A\sim\frac{c\pi T t}{8}.
\end{align}
Therefore the second R\'enyi entropy grows linearly, again indicating the exponential deviation of the OTOC.

\section{The Lyapunov Exponent} 
\label{lya}

\begin{figure*}[t]
\includegraphics[width=\textwidth]{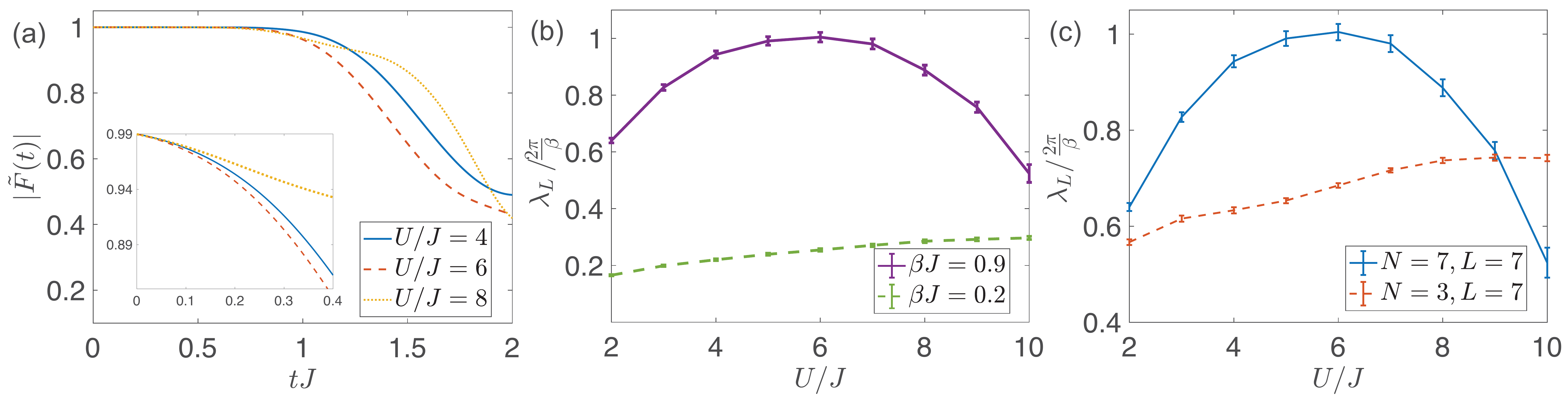}
\caption{(a) The amplitude of normalized OTOC $|\tilde{F}(t)|$ as a function of time $tJ$ for $U/J=4, 6$ and $8$ at $\beta J=0.9$ and $N=L=7$. $ N $ is the number of bosons and $ L $ is the system size. The inset is a zoom-in plot of the early-time deviation behavior with $ t_0 $ aligned together. It is clear that the $ U/J=6 $ curve deviates faster than the $ U/J=4 $ and $ 8 $ curves. (b-c) The Lyapunov exponents as a function of $U/J$. The error bars come from the fitting. (b) is plotted for $\beta J=0.9$ and $ 0.2$ with $ N=L=7 $; (c) is plotted for $N=7$ and $N=3$ with $L=7$, $\beta J=0.9$. In all the three figures above, we have chosen $\hat{V}=\hat{b}_1$, $\hat{W}=\hat{b}_4$ and the periodic boundary condition. For the fitting, we take the fitting parameters $F_\mathrm{c}=0.99 $ and $p=0.2$. We have verified that changing the fitting parameters will not affect the trend of the data, but will only modify the exponents quantitatively. }
\label{OTOCbi}
\end{figure*}

Having shown that the OTOC should deviate exponentially in time, we
are now in a position to extract the Lyapunov exponent. Three typical curves of the OTOC are shown in Fig.~\ref{OTOCbi}(a). In order to fit the Lyapunov exponent at the early time, we adapt the following fitting scheme shown in Fig.~\ref{schematic}(b):

\begin{enumerate}
  \item We choose a threshold $F_\text{c}$ ($F_\text{c}\lesssim 1$) to determine a starting time $t_0$ as $\tilde{F}(t_0)=F_\text{c}$. $t_0$ is the initial time when the OTOC starts to deviate exponentially. 
  
  \item The second-order derivative of $\tilde{F}(t)$ is denoted as $\tilde{F}''(t)$. We take $t_2$ to be the last point (after $t_0$) that satisfies $\tilde{F}''(t)<0$. In other words, for $t>t_2$, $\tilde{F}''(t)>0$ and obviously $\tilde{F}(t)$ can no longer be fitted by an exponential.
  
  \item In fact, the OTOC deviates from the exponential even before reaching $t_2$. Therefore we introduce another parameter $p$, which we call the ``retaining fraction''. Assuming all data points are uniformly taken along the time direction, we define $t_1<t_2$ to satisfy $(t_1-t_0)/(t_2-t_0)=p$. The principle of choosing $ p $ is to set $ p $ as large as possible as long as the error of the fitting is small.
  
  \item We fit all the data points between $t_0 $ and $t_1$ by a function $f(t)=Ae^{\lambda_\mathrm{L}t}+B$. We take the logarithm of the first-order derivative of $f(t)$ as
  \begin{equation}
  \log(f'(t)) = \log(A\lambda_\mathrm{L}e^{\lambda_\mathrm{L}t})= \log(A\lambda_\mathrm{L}) + \lambda_\mathrm{L}t ,
  \end{equation}
  where the Lyapunov exponent $\lambda_\mathrm{L}$ is just the slope of this linear regression $ \log(f'(t))\sim t $. 
\end{enumerate}

Before presenting our results, we would like to comment on the separation of time scales in our calculation. There are two time scales involved: the dissipation time $t_\mathrm{d}$ and the scrambling time $t_\mathrm{s}$ \cite{Maldacena2016}, which can be extracted from the normal time-order correlators and the OTOC respectively. Roughly speaking, $t_\mathrm{d}$ characterizes the time when the excitation $\hat{V}(0)|\beta\rangle$ is smeared out, so the normal time-ordered correlator factorizes as $\langle \hat{V}^\dag(0)\hat{W}^\dag(t)\hat{W}(t)\hat{V}(0)\rangle=\langle\hat{V}^\dag\hat{V}\rangle\langle\hat{W}^\dag\hat{W}\rangle$. The scrambling time $t_\mathrm{s}$ characterizes the time when the information is scrambled and is identified when $\tilde{F}(t)$ first reaches its local minimum. In order for the scrambling to be well-defined, the separation of time scale is required, i.e., the scrambling takes place at $t_\mathrm{d}\ll t < t_\mathrm{s}$. This usually requires a large number of degrees of freedom such as those in some large-$ N $ models.

Here, we consider the case when $\hat W$ and $\hat V$ are located at different sites so that their spatial distance can guarantee the separation of time scale. For operators with spatial separation $ |x| $, the OTOC could be expanded as
\begin{equation}
\tilde{F}(t)=\alpha_0-\alpha_1 e^{\lambda_{\mathrm{L}}(t-|x|/v_{\mathrm{B}})}, \label{bf}
\end{equation}
where the small parameter $ e^{-\lambda_{\mathrm{L}}|x|/v_{\mathrm{B}}} $ suppresses the high order terms in the expansion. $v_{\mathrm{B}}$ is called the ``butterfly velocity'' \cite{Roberts2015,Roberts2016}, which is to be discussed in detail in the next section.

We plot three OTOCs at temperature $\beta J=0.9$ starting from their corresponding $t_0$ in the inset of Fig.~\ref{OTOCbi}(a). As can be seen clearly, the deviation first becomes more rapid as $U/J$ increases (from 4 to 6), and then becomes slower as $U/J$ further increases (from 6 to 8). By fitting the Lyapunov exponents using the scheme introduced above, we find that $\lambda_\mathrm{L}$ displays a broad peak around $U/J=6$, and the peak value is very close to the $ 2\pi/\beta $ bound (Fig.~\ref{OTOCbi}(b)). It is instructive to consider the system at high temperature that is away from the quantum critical region. For temperature as high as $\beta J=0.2$, not only the peak of the Lyapunov exponent disappears, but the magnitude of these exponents are considerably smaller compared with the bound. 

To further confirm that the peak indeed comes from the quantum criticality, we calculate the OTOC in the system that is away from the integer filling and hence the quantum critical region. As shown in Fig.~\ref{OTOCbi}(c), there is no peak in $\lambda_\mathrm{L}$ as $U/J$ increases. Also, the Lyapunov exponents are generally smaller compared to those of the systems at integer filling under the same temperature.

Before proceeding, we would like to make several remarks on the finite-size effect. At even lower temperatures, we find that $\lambda_\mathrm{L}$ could exceed the $ 2\pi/\beta $ bound. We attribute this behavior to two reasons. First, the low temperature is only well-defined when the temperature is still higher than the finite-size gap. While in the BHM of size $ L=7 $, the finite-size gap is comparable to $ \beta J \sim 1 $. So $ \beta J=0.9 $ is almost the lowest temperature we could reach in order to obtain reliable results. Second, the proof of the bound on chaos relies heavily on the large hierarchy between the dissipation time $ t_{\mathrm{d}} $ and the scrambling time $ t_{\mathrm{s}} $ \cite{Maldacena2016}, which may be missing in a system of very limited size due to the numerics \cite{Fu2016,Yao2016}. Also, for $ (1+1) $-dimensional BHM, the zero-temperature quantum critical point is located at $U/J\sim 3.4$ \cite{Kuhner1999}, while the peak of $\lambda_\mathrm{L}$ appears at $U/J\sim 6$ in our calculation. This discrepancy may be due to the fact that our calculation is done at a still relatively high temperature as $\beta J=0.9$, where the quantum critical region already spans a quite broad area in the parameter space. 

\begin{figure}
\includegraphics[width=\columnwidth]{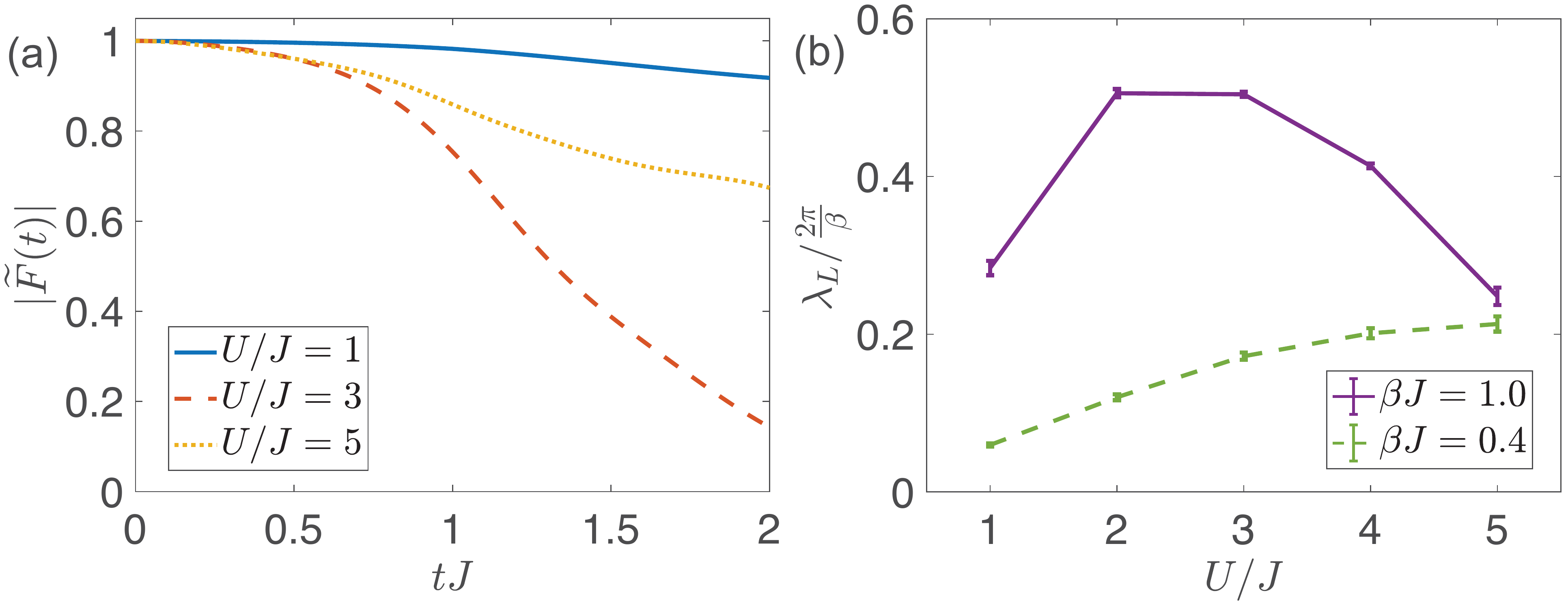}
\caption{(a) The amplitude of normalized OTOC $|\tilde{F}(t)|$ as a function of time $tJ$ for $U/J=1,3$ and $5$ at $\beta J=1.0$. (b) The Lyapunov exponents as a function of $U/J$ plotted for $\beta J=1.0$ and $0.4$. The error bars come from the fitting. In all the two figures above, we have chosen $ N=L=6 $ with a periodic boundary condition, $\hat{V}=\hat{W}=\hat{b}_{\bf k}$, $ k=\pi/3 $. For the fitting, we take the fitting parameters $F_\mathrm{c}=0.99 $ and $p=0.8$. We have verified that changing the fitting parameters will not affect the trend of the data, but will only modify the exponents quantitatively. }
\label{OTOCbk}
\end{figure}

We end this section by considering the dependence of the Lyapunov exponent $\lambda_\mathrm{L}$ on the choice of the operators. We rewrite the BHM into momentum space as
\begin{equation}
\hat{H}=\sum\limits_{{\bf k}}\epsilon_{{\bf k}}\hat{b}^\dag_{{\bf k}}\hat{b}_{{\bf k}}+\frac{U}{2L}\sum\limits_{{\bf k}_1{\bf k}_2{\bf k}_3{\bf k}_4}\hat{b}^\dag_{{\bf k}_1}\hat{b}^\dag_{{\bf k}_2}\hat{b}_{{\bf k}_3}\hat{b}_{{\bf k}_4}, \label{bhmk}
\end{equation}
where $\epsilon_{{\bf k}}=2J\cos k-U/2$ and ${\bf k}_1+{\bf k}_2={\bf k}_3+{\bf k}_4$. Instead of the real space boson operators, we can choose the momentum space boson operators $\hat{b}_{\bf k}$ as $\hat{V}$ and $\hat{W}$. In this way, $\hat{b}_{{\bf k}}$ can be regarded as local operators in the momentum space, although the model Eq.~\eqref{bhmk} now has infinite range interactions. We also find a peak in the Lyapunov exponent as $ U/J $ varies, as shown in Fig.~\ref{OTOCbk}. The peak of $\lambda_\mathrm{L}$ is closer to the zero-temperature critical point in this case.

\section{The Butterfly Velocity} 
\label{but}
\begin{figure}[t]
\includegraphics[width=\columnwidth]{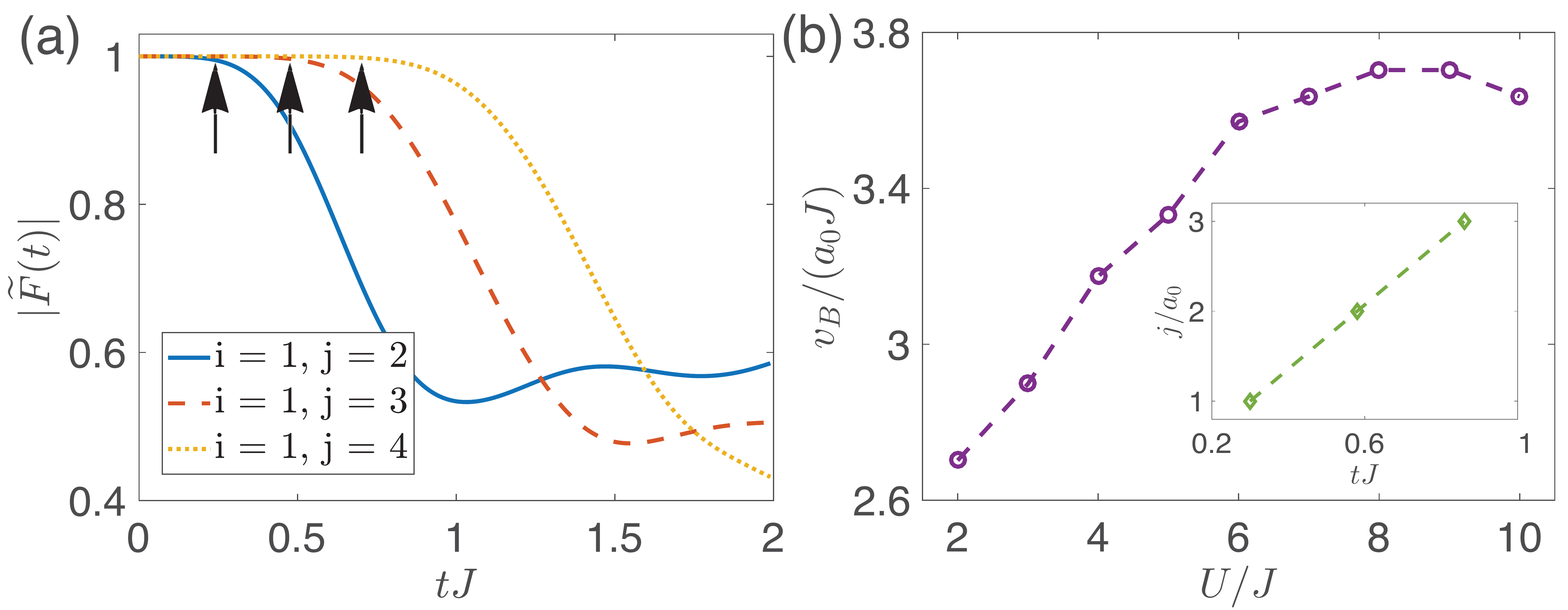}
\caption{(a) The amplitude of normalized OTOC $|\tilde{F}(t)|$ as a function of time $ tJ $ for $ U/J=6 $. $\hat{V}=\hat{b}_i$ and $\hat{W}=\hat{b}_j$ with $i$ fixed at $i=1$ and $j$ varies between $j=2$, $j=3$ and $j=4$. (b) The butterfly velocity extracted from the OTOC. $ a_0 $ is the lattice spacing. The inset is the time $ t_0 $ where the OTOC begins to deviate exponentially as a function of the site $ j $ for $ U/J=6 $. In all the two plots above, $\beta J=0.9$ and $N=L=7$ with periodic boundary condition. To extract $ t_0 $ we choose $ F_{\mathrm{c}} =0.99$. }
\label{Bv}
\end{figure}

Now we turn to the discussion of the butterfly velocity $ v_{\mathrm{B}} $ appeared in Eq.~\eqref{bf}. It is defined in systems with spatial degree of freedom and describes how fast the information propagates along the spatial directions. We consider $\hat{V}=\hat{b}_i$ and $\hat{W}=\hat{b}_j$, where $i$ and $j$ are at different sites. $t_0$, previously defined as the time for the onset of the deviation, increases linearly with the distance between $i$ and $j$, as shown in the Fig.~\ref{Bv}(a) and the inset of Fig.~\ref{Bv}(b). From this slope we can extract the butterfly velocity and the results are shown in Fig.~\ref{Bv}(b). We find that the butterfly velocity first increases with $U/J$. At large $U/J$ it seems to saturate and even begin to decrease weakly. 

It is interesting to compare the butterfly velocity with the Lieb-Robinson velocity \cite{Lieb1972,Hastings2006,Nachtergaele2006}, which has been studied both numerically \cite{Laeuchli2008} and experimentally \cite{Cheneau2011} for the BHM. Since the Lieb-Robinson velocity can roughly be regarded as the butterfly velocity at infinity temperature, intuitively they should share the same trend but the butterfly velocity is smaller. This is indeed what we find in our calculation.

\section{Experiment Protocol of Measuring OTOC} 
\label{expp}
Finally, we discuss the experiment protocol of measuring OTOCs. So far, all the existing proposals for measuring OTOCs rely on the ability to evolve the system backward in time \cite{Swingle2016,Zhu2016,Yao2016}, i.e. to invert the entire Hamiltonian from $ \hat{H} $ to $ -\hat{H} $. We first remark that this is also doable for BHM in cold atom realizations \cite{Jaksch2005,Bloch2008,Yukalov2009}. To invert $U$, one can use the Feshbach resonance to change the sign of the $s$-wave scattering length. To invert hopping, one can exploit the technique of shaking optical lattice. According to the Floquet theory, the hopping in a periodically shaking lattice is modified by the zeroth-order Bessel function as $JJ_0(Aa_0m\omega)$, where $a_0$ is the original lattice spacing, $A$ is the shaking amplitude, $\omega$ is the shaking frequency and $m$ is the mass of atoms. Thus, one can tune the shaking frequency from $ \omega_1 $ to $ \omega_2 $ such that $J_0(Aa_0 m\omega_1)=-J_0(Aa_0 m\omega_2)$. Moreover, there is no intrinsic difficulty that prevents performing these operations simultaneously. Therefore, the total Hamiltonian is inverted. 

Here we propose an alternative way to measure OTOC, which does not require inverting the Hamiltonian. Instead, it demands preparing two identical copies of the system. The spirit is similar to the recent measurements of the second R\'enyi entropy in the BHM using a Hong-Ou-Mandel-type interference \cite{Hong1987,Daley2012,Islam2015}. The modified OTOC \cite{Maldacena2016} to be measured is 
\begin{align}
F_M(t)=\mathrm{tr}\left[\hat{W}^\dagger(t)\hat{O}e^{-\beta \hat{H}/2}\hat{O}^\dagger \hat{W}(t)\hat{O}e^{-\beta \hat{H}/2}\hat{O}^\dagger \right].
\end{align}
Similar to the discussion in Sec.~\ref{expo}, here $\hat{V}=\hat{O}\hat{O}^\dagger$. Since $\hat{W}(t)=e^{i\hat{H}t}\hat{W}e^{-i\hat{H}t}$, 
\begin{align}
&F_M(t)\notag\\&=\mathrm{tr}\left[e^{i\hat{H}t}\hat{W}^\dagger e^{-i\hat{H}t}\hat{O}e^{-\beta \hat{H}/2}\hat{O}^\dagger e^{i\hat{H}t}\hat{W}e^{-i\hat{H}t}\hat{O}e^{-\beta \hat{H}/2}\hat{O}^\dagger \right]\notag\\
&=\mathrm{tr}\left[\hat{\rho}_1\hat{\rho}_2\right]=\mathrm{tr}\left[\hat{S}_{12}\hat{\rho}_1\otimes\hat{\rho}_2\right],
\end{align} 
where 
\begin{align}
&\hat{\rho}_1=\hat{W}^\dagger e^{-i\hat{H}t}\hat{O}e^{-\beta \hat{H}/2}\hat{O}^\dagger e^{i\hat{H}t}\hat{W},\\
&\hat{\rho}_2=e^{-i\hat{H}t}\hat{O}e^{-\beta \hat{H}/2}\hat{O}^\dagger e^{i\hat{H}t}.
\end{align} 
The normalization is $\mathrm{tr}\left[\hat{\rho}_i\right]=1$. $\hat{S}_{12}$ is the swap operator that exchanges states in the two copies of the system $ \hat{S}_{12}|{\psi_i}\rangle\otimes|{\psi_j}\rangle=|{\psi_j}\rangle\otimes|{\psi_i}\rangle $. In this way, the modified OTOC is reformulated into the interference of two density matrices $\mathrm{tr}[\hat{\rho}_1\hat{\rho}_2]$, and can be measured using the same protocol described in Ref.~\cite{Daley2012,Islam2015}. 

In summary, the experiment protocol for measuring the modified OTOC between $\hat{W}$ and $\hat{V}=\hat{O}\hat{O}^\dagger$ at temperature $T$ is as follows: 
\begin{enumerate}
  \item Prepare two identical copies of the systems at temperature $2T$;
  \item Suddenly quench both systems by applying operator $\hat{O}$ on both copies;
  \item Let both copies evolve under the Hamiltonian $\hat{H}$ for a duration of time $t$;
  \item Apply the operator $\hat{W}$ to only one of the copies;
  \item Perform a Hong-Ou-Mendel-type interference of the two systems.
\end{enumerate}

We note that this scheme is closely related to the Loschmidt echo experiment, which has recently been found closely related to the OTOC \cite{Kurchan2016}. Having been performed in many quantum systems, the Loschmidt echo experiments may shed light on future studies of the OTOC. 

\section{Remarks and the Outlook}

Despite the holographic duality argument given in the introduction, there is also an intuitive argument to understand the peak in the Lyapunov exponent. For $U=0$, the Hamiltonian describes non-interacting bosons in a lattice. As $U$ increases, the interaction effect gradually raises $\lambda_\mathrm{L}$. On the other hand, in the large-$U$ limit, the Hamiltonian and all commutators can be expanded perturbatively in terms of $ J/U $. At the zeroth order $ J/U=0 $, each site becomes independent and the OTOC does not change with time. The Lyapunov exponent should increase as $ J/U $ decreases. Thus we would expect that $\lambda_\mathrm{L}$ has a peak in between. 

In fact, the underlying insight from the condensed matter physics is that there are no well-defined quasiparticles in the strongly interacting quantum critical region. Therefore, the system is more chaotic than that in the non-critical region. As a result, the Lyapunov exponent should be larger in the quantum critical region. For example, we have also studied the quantum phase transition in the XXZ model and the transverse field Ising model, where similar phenomena are found. For the XXZ model $ \hat{H}=-J_\perp\sum_{i} (\hat{s}^x_i \hat{s}^x_{i+1}+\hat{s}^y_i \hat{s}^y_{i+1})-J_z \sum_{i}\hat{s}^z_i \hat{s}^z_{i+1} $, where $ \hat{s}^\alpha_{i}  $, $ \alpha=x,y,z $ are spin operators at $ i $-th site, we choose $\hat{W} $ and $\hat{V}$  as $\hat{s}^+_{i}-\hat{s}^+_{i+1} $ at different sites, whose bosonization representation is the same as that of $\hat{b}^\dag_i$ in BHM. For the transverse field Ising model $ \hat{H}=-J\sum_{i}\hat{s}^z_{i}\hat{s}^z_{i+1}-g\sum_{i}\hat{s}^x_{i} $, we use the open boundary condition and choose boundary operators $ \hat{s}^+_{1} $ and $ \hat{s}^+_{L} $ to characterize the phase transition. In both cases, we find a broad peak of the Lyapunov exponent around the quantum critical region. 

Therefore, we believe that our \textit{QCP Conjecture} for the Lyapunov exponent is very general. This conjecture could be tested by more theoretical and experimental studies in the future. 

\textit{Acknowledgment.} We would like to thank Yingfei Gu and Chao-Ming Jian for helpful discussions. This work is supported by MOST under Grant No. 2016YFA0301600, NSFC Grant No. 11325418 and Tsinghua University Initiative Scientific Research Program.

\appendix
\section{CFT Calculation of The R\'enyi Entropy Growth after a Local Quench}

\begin{figure}[t]
	\includegraphics[width=\columnwidth]{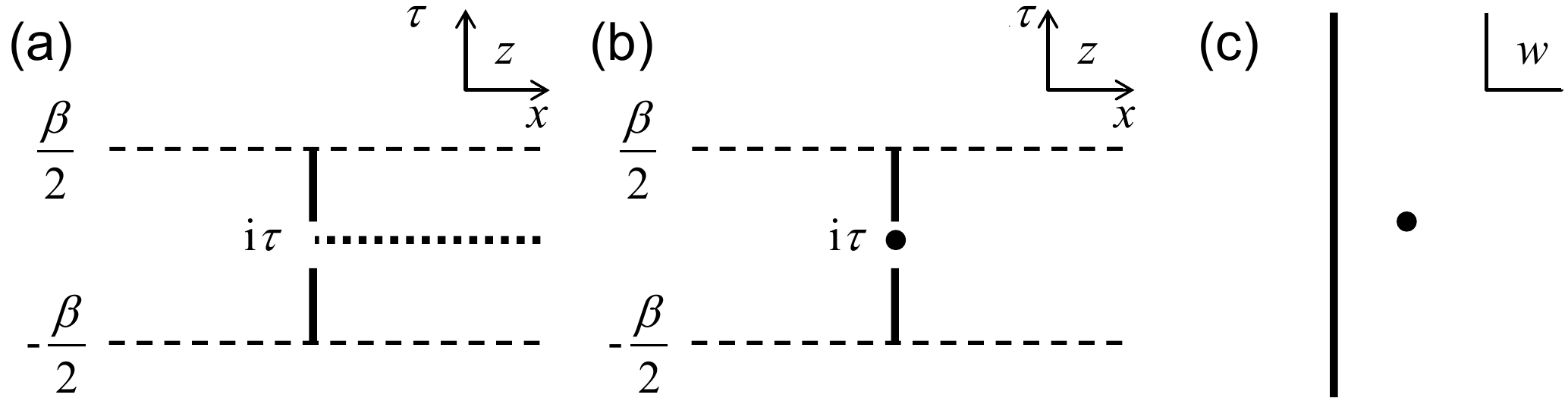}
	\caption{The procedure of the conformal field theory calculation. (a) The original geometry on the stripe. There is a physical boundary along the imaginary axis and a cut along $+x$ direction from $z=i\tau_0$. For tr$\left[\rho^n\right]$ we need $n$ copies and sew them together. (b) The complicated Riemann surface is identified with a twist field at $z=i\tau_0$ with $n$ copies of field on a single stripe. (c) After the conformal mapping, the problem becomes a standard geometry for a half infinite plane.}
	\label{cft}
\end{figure}

In this appendix, we derive Eq.~\eqref{cftres} in detail. The technique used here is similar to that in Ref.~\cite{Calabrese2009}. The main difference is that our CFT is defined on a stripe because our system is at finite temperature, while theirs is defined on the full plane due to the zero temperature. 
 
The system is put on on a stripe with a periodic boundary along the imaginary time direction as shown in Fig.~\ref{cft}(a), and is divided as part $ A $ and $ B $ for $ x>0 $ and $ x<0 $ respectively. The evolution of each part is governed by $\hat{H}'$, and there is no interaction between them. The local quench is achieved by connecting $ A $ and $ B $ in a small time window $\epsilon$ near $i\tau_0$. We put a cut along $ +x $ space direction at time $\tau_0$ because there is no trace over subsystem $ A $, after which the evolution of the whole system is governed by $ \hat{H} $. The value of the field at the upper (lower) branch of the cut-line gives the row (column) index of the reduced density matrix $\hat{\rho}_A$:
\begin{align}
\hat{\rho}_\mathrm{A}=\mathrm{tr}_B\left[\exp(-\hat{H}(\epsilon+\tau_0))\exp(-\beta \hat{H}^\prime)\exp(-\hat{H}(\epsilon-\tau_0)) \right]. \label{rho}
\end{align}

The calculation of $n$-th R\'enyi entropy requires $n$ copies of the stripe. We use the complex coordinate $z=x+i\tau$ for each stripe and sew these stripes one by one by imposing the boundary condition $\phi_{i+1}(x+i\tau_0+i0^+)=\phi_{i}(x+i\tau_0-i0^+)$ for $x>0$. Here the label of the copies $ i $ is defined mod $ n $. $\phi_i$ is the field in the CFT on the $ i $-th copy of the stripe, and is also the label of the coherent state for corresponding operator. As a result, the sewed stripes form a complicated Riemann surface. Then we have a functional integral with the boundary constraint
\begin{equation}
\mathrm{tr}\left[\hat{\rho}_A^{n}\right]=\int_{\mathrm{bound. con.}}\mathcal{D}\phi_i \exp\left(-\sum_iS[\phi_i]\right), 
\end{equation}
where $S[\phi]$ is the action for a single copy of the field $\phi$. Equivalently, the boundary constraint can be replaced by introducing a twist field $\mathcal{T}_n(i\tau_0)$ that acts at time $\tau_0$ and swaps the value of field to the right of $x$:
\begin{equation}
\mathrm{tr}\left[\hat{\rho}_A^{n}\right]=\int\mathcal{D}\phi_i \mathcal{T}_n(i\tau_0)\exp\left(-\sum_iS[\phi_i]\right).
\end{equation}
$\mathcal{T}_n$ is known as a primary field with conformal dimension $d_n=\frac{c}{12}\left(n-\frac{1}{n}\right)$ where $c$ is the central charge. The configuration is shown in Fig. \ref{cft}(b). 

This strip ($z$) can be mapped to a half infinite plane with a boundary at imaginary axis ($w$) using conformal transformation: 
\begin{equation}
\epsilon w=\tanh(\pi z/\beta)+\sqrt{\tanh(\pi z/\beta)^2+\epsilon^2}.
\end{equation}
The standard formula for the conformal field theory with a boundary \cite{Kurchan2016} gives the expectation on the stripe shown in Fig.~\ref{cft}(c):
\begin{equation}
\mathrm{tr}[\rho_A^n]=\left<\mathcal{T}_n(z)\right>=c_n\left(\left|\frac{dw}{dz}\right|\frac{1}{2 \mathrm{Re}\,w}\right)^{d_n},
\end{equation}
where $c_n$ is some constant. After the analytical continuation back to the real time and taking cutoff $\epsilon$ to zero, we obtain the result Eq.~\eqref{cftres}.

\bibliography{QPT.bib}

\end{document}